\documentclass[pra,epsfig,rotate,superscriptaddress,showpacs,11pt,tightenlines]{revtex4}
\usepackage[pdftex]{graphicx}
\usepackage{amsfonts}
\usepackage{amssymb}
\usepackage{amsmath}
\usepackage{latexsym} 
\usepackage{bm}   
\DeclareGraphicsExtensions{.pdf,.png,.jpg}
\newcommand{\p}{\partial}
\newcommand{\<}{\langle}
\renewcommand{\>}{\rangle} 
\newcommand{\txt}{\textstyle}

\newcommand\eqn[1]{(\ref{#1})}      
\newcommand{\e}{{\rm e}}   
\newcommand{\ri}{{\rm i}}
\newcommand{\rd}{{\rm d}}
\newcommand{\one}{\openone}
\newcommand{\beq}{\begin{equation}}
\newcommand{\eeq}{\end{equation}}
\newcommand{\bea}{\begin{eqnarray}}
\newcommand{\eea}{\end{eqnarray}}

\newcommand{\half} {{\txt \frac{1}{2}}}

\newcommand{\cP}{\ensuremath{\mathcal{P}}}
\newcommand{\cT}{\ensuremath{\mathcal{T}}}
\newcommand{\ie}{{i.e.}}
\newcommand{\eg}{{e.g.}}
\newcommand{\nn}{\nonumber \\}

\begin{document}

\title{\bf Time-dependent $\mathcal{PT}$-symmetric quantum mechanics}

\author{Jiangbin Gong}
\affiliation{Department of Physics, National University of Singapore, 117542,
Singapore}
\affiliation{Centre for Computational Science and Engineering, National University of Singapore, 117542, Singapore}
\affiliation{NUS Graduate School for Integrative Sciences and Engineering, 117597, Singapore}

\author{Qing-hai Wang}
\affiliation{Department of Physics, National University of Singapore, 117542,
Singapore}

\date{November 1, 2013}

\begin{abstract}
  The parity-time-reversal- ($\mathcal{PT}$) symmetric quantum mechanics (PTQM) has developed into a noteworthy area of research. However, to date most known studies of PTQM focused on the spectral properties of non-Hermitian Hamiltonian operators.  In this work, we propose an axiom in PTQM in order to study general time-dependent problems in PTQM, e.g., those with a time-dependent $\mathcal{PT}$-symmetric Hamiltonian and with a time-dependent metric.  We illuminate our proposal by examining a proper mapping from a time-dependent Schr\"odinger-like equation of motion for PTQM to the familiar time-dependent Schr\"odinger equation in conventional quantum mechanics.  The rich structure of the proper mapping hints that time-dependent PTQM can be a fruitful extension of conventional quantum mechanics.  Under our proposed framework, we further study in detail the Berry phase generation in a class of $\mathcal{PT}$-symmetric two-level systems.  It is found that a closed path in the parameter space of PTQM is often associated with an open path in a properly mapped problem in conventional quantum mechanics. In one interesting case we further interpret the Berry phase as the flux of a continuously tunable fictitious magnetic monopole, thus highlighting the difference between PTQM and conventional quantum mechanics despite the existence of a proper mapping between them.
\end{abstract}
\begin{titlepage}
\maketitle
\end{titlepage}

\section{Introduction}
As a fruitful extension of conventional quantum mechanics (QM), the parity-time-reversal- (\cP\cT) symmetric quantum mechanics (PTQM) has developed into a noteworthy area of research \cite{BB,BenderReview}.  So far, the main emphasis of PTQM  studies has been placed, both theoretically and experimentally, on the spectral properties of non-Hermitian Hamiltonians.  Due to a nonconventional and well-behaved inner-product structure, the spectrum of a complex but \cP\cT-symmetric non-Hermitian Hamiltonian is still real if the \cP\cT\ symmetry is not spontaneously broken.  This fascinating result hints the potential of introducing a non-trivial inner-product structure to QM \cite{BBJ}.

In conventional QM,  the spectral problem for a Hamiltonian or energy operator is tackled by the stationary Schr\"odinger equation. The general equation of motion is given by a second equation, i.e., the time-dependent Schr\"odinger equation that describes how a quantum system evolves with time. There are opposite views about which of the two Schr\"odinger equations is more vital in conventional QM. In one view, the stationary Schr\"odinger equation can be regarded as a practical tool to solve the time-dependent Schr\"odinger equation, with the latter being a postulate in QM.   Indeed, Dirac directly put down the time-dependent Schr\"odinger equation with an arbitrary time-dependent Hamiltonian, by reasoning that the Hamiltonian or the energy operator of a system should be a generating operator of time displacements \cite{Dirac}.

Given the importance of the time-dependent Schr\"odinger equation in conventional QM, it is necessary to seek a general equation of motion in PTQM. In some early studies of time-dependent PTQM, the conventional Schr\"odinger equation was used without being questioned \cite{Fring,Mostafazadeh}. In these cases, a notable constraint is that although the Hamiltonian is time-dependent, the inner-product remains to be stationary in order to have unitary evolution. However, this constraint may be lifted because a time-dependent inner-product and the unitary condition can be made compatible with a time-dependent Schr\"odinger-{\it like} equation \cite{Znojil,GW}. That is, to study time-dependent PTQM with a time-dependent inner-product, one has to go beyond the conventional time-dependent Schr\"odinger equation. The first attempt to construct a Schr\"odinger-like equation for PTQM was made in \cite{Znojil} by mapping a Hermitian system to a non-Hermitian \cP\cT-symmetric one using a known mapping. In this sense, the evolution of a time-dependent PTQM system is generated by an already-known Hermitian system. Taking a different approach, in our previous work \cite{GW} we attempted to formulate PTQM as a fundamental theory and treat conventional QM as a special case of PTQM. However, there it was noted that for PTQM, there is ambiguity when constructing a time-dependent Schr\"odinger-like equation yielding unitary evolution \cite{GW}. That is, given a time-dependent \cP\cT-symmetric Hamiltonian with a time-dependent inner-product, there can exist an infinite number of choices for the time evolution operator yielding a unitary evolution.  In retrospect, this must be the case because even in conventional QM, the unitarity requirement does not suffice to determine the form of a time-dependent Schr\"odinger equation.
The explicit results reported in \cite{GW} were based on a particular type of  Schr\"odinger-like equation, without much justification.

In this work we seek an equation of motion for PTQM that is as general as possible, so as to cover cases with (i) time-dependent non-Hermitian Hamiltonians and (ii) time-dependent inner-product structure. Such efforts devoted to time evolution problems in PTQM will supplement ongoing PTQM activities on the spectral side.  Time evolution in PTQM is also a necessary subject in order to treat conventional QM as a special case of PTQM.  In achieving our goal we propose an axiom to remove the ambiguity mentioned above.  To that end we first note a dual role of Hamiltonian operators in conventional QM and then make a close analogy in PTQM.  Further, we illuminate our axiom by examining the mapping of equation of motion from PTQM to that in conventional QM.  We are able to identify a special type of mapping, termed as ``proper mapping'' below, from which we gain interesting insights into time-dependent PTQM.  We use the complex harmonic oscillator problem as a simple working example to illustrate a proper mapping.

To further elaborate our theoretical proposal, we revisit the Berry-phase problem for a class of \cP\cT-symmetric two-level systems \cite{GW}. The results can be associated with a fictitious nontrivial Dirac string and a fictitious continuously tunable monopole. To explain these counter-intuitive results found in \cite{GW}, we investigate a proper mapping between conventional QM and PTQM.  In particular, it is found that a cyclic parameter path in PTQM is not necessarily a cyclic one with the same period for a properly mapped problem in conventional QM.  This explanation is one important insight offered by our axiom proposed in this work.  It is also our belief that these specific results should stimulate further general interests in time-dependent PTQM, with our proposed axiom as a promising starting point.

The organization of this paper is as follows. In Section \ref{sec:EoM}, we first make remarks on the time-dependent Schr\"odinger equation in conventional QM and then propose a Schr\"odinger-like equation of motion for PTQM.  We further define a proper mapping between these two theories. In Section \ref{sec:CHO} we adopt a simple example to illustrate our time-dependent PTQM.  The Berry phase in a class of \cP\cT-symmetric two-level systems are presented and discussed in depth in Section \ref{sec:Berry}. Section \ref{sec:conclusions} concludes this work. Some helpful derivations are also presented in Appendices.

\section{Schr\"odinger equation of motion: from conventional QM to PTQM}
\label{sec:EoM}

\subsection{Remarks on Hamiltonian operators in conventional QM}
In conventional quantum mechanics, the Hamiltonian operator $h$ determines the spectrum of a system through the stationary Schr\"odinger equation,
\begin{equation}
h |\phi_n\> = E_n |\phi_n\>,
\label{eqn:schr}
\end{equation}
where $E_n$ are the eigenvalues and $|\phi_n\>$ represent the eigenkets. Even if $h$ is time-dependent, the above eigenvalue equation is still useful as it yields an instantaneous spectrum for the system. (The instantaneous spectrum also gives the possible measurement results if the energy value is measured.)  Note that upon a unitary transformation $U$, the above equation is invariant with $h\rightarrow h'=UhU^{\dagger}$ and $|\phi_n\> \rightarrow |\phi'_n\>=U|\phi_n\>$.

The time evolution of the above system is governed by the following time-dependent Schr\"odinger equation,
\begin{equation}
\ri \hbar \frac{\rd}{\rd t} |\Phi(t)\> = h(t) |\Phi(t)\>,
\label{eqn:full_schr}
\end{equation}
where we have used $h(t)$ to stress that the Hamiltonian operator $h$ can be explicitly time-dependent. Dirac justified this equation of motion by noting that, in this form the energy operator becomes the generator of time displacements, which is then analogous to the conjugate pair of position and momentum variables \cite{Dirac}. However, such an equation of motion, which is a postulate in conventional QM, is only invariant under a time-independent unitary transformation. It is {\it not}\ invariant under arbitrary time-dependent unitary (gauge) transformations. Indeed, if we apply a time-dependent unitary transformation, $h(t)\rightarrow h'(t)=U(t)h(t)U^\dag(t)$ and $|\Phi(t)\> \rightarrow |\Phi'(t)\>=U(t)|\Phi(t)\>$, then (\ref{eqn:full_schr}) is known to become
\begin{equation}
\ri \hbar \frac{\rd}{\rd t} |\Phi'(t)\> = \left[h'(t) - \ri\hbar U(t)\dot{U}^\dag(t)\right] |\Phi'(t)\>,
\label{eqn:SchrU}
\end{equation}
where an overhead dot represents the time derivative. As a result, some appropriate representation (gauge choice) must be implicitly adopted first when writing down the time-dependent Schr\"odinger equation (\ref{eqn:full_schr}). For our later use,  we stress the evident observation here: the Hamiltonian or the energy operator in conventional QM plays a dual role: the operator that determines the energy spectrum of the system is also taken as the generator of time displacements.  Later we shall exploit this dual role as a basis to define a ``proper" mapping between conventional QM and PTQM.

\subsection{Time-dependent \cP\cT-symmetric quantum mechanics}
As is well known now, the Hilbert space in PTQM is associated with a metric operator $W$ that defines an inner product \cite{BBJ}. We consider PTQM in its most general form by studying time-dependent Hamiltonian operators $H(t)$ and also time-dependent metric operator $W(t)$ associated with $H(t)$. At any instant $t$ under investigation, $H(t)$ is assumed to have an extended \cP\cT\ symmetry characterized by
\begin{equation}
W(t)H(t)=H(t)^\dag W(t).
\label{eqn:metric}
\end{equation}
The inner product between bra and ket states becomes $\langle \cdot |W|\cdot \rangle$, where a bra state is defined as the Dirac conjugate of a ket state, $\<\cdot| \equiv |\cdot\>^\dag$. Note that we do not need to introduce a bi-orthonormal basis as commonly used in the study of open quantum systems with non-Hermitian Hamiltonians. The reason is that  our \cP\cT-symmetric Hamiltonian is already assumed to be self-adjoint (under the metric $W$) when \cP\cT\ symmetry is unbroken. For a self-adjoint operator $H(t)$, all its eigenvalues are real and its instantaneous eigenstates in
\begin{equation}
H(t)|\psi_n(t)\> = E_n(t)|\psi_n(t)\>
\label{eqn:PTeigen}
\end{equation}
form a set of orthonormal basis with respect to the metric $W(t)$,
\begin{equation}
\<\psi_m(t)|W(t)|\psi_n(t)\> = \delta_{mn}.
\label{eqn:orthonormal}
\end{equation}
To construct an equation of motion, we impose the following unitary condition
\begin{equation}
\frac{\rd}{\rd t}\<\Psi_1(t)|W(t)|\Psi_2(t)\>=0,
\end{equation}
where $|\Psi_1(t)\>$ and $|\Psi_2(t)\>$ are two time evolving states from two different initial conditions $|\Psi_1(0)\>$ and $|\Psi_2(0)\>$. As already shown in our previous work~\cite{GW}, to satisfy the above unitary condition the equation of motion must take the following form
\begin{equation}
\ri\hbar \frac{\rd}{\rd t}|\Psi(t)\> = \Lambda (t) |\Psi(t)\>,
\end{equation}
where the time-displacement generator $\Lambda$ satisfies (from now on, we sometimes suppress the time argument)
\begin{equation}
\ri \hbar \dot{W} = \Lambda^\dag W - W \Lambda.
\label{eqn:Wdot}
\end{equation}
Interestingly, $\Lambda$ must {\it not}\ equal $H$ when the metric operator is changing in time [because $WH-H^{\dagger}W=0$ according to (\ref{eqn:metric})].  This hints that general time-dependent problems in PTQM go far beyond the spectral properties of a \cP\cT-symmetric Hamiltonian operator.

Just like the fact that a generic matrix can be written as the sum of a Hermitian matrix and an anti-Hermitian matrix, we may also divide $\Lambda$ into two parts,
\begin{equation}
\Lambda = \tilde{H} + A,
\label{eqn:part}
\end{equation}
with
\begin{equation}
W\tilde{H} = \tilde{H}^\dag W,\qquad {\rm and} \qquad
W A = -A^\dag W.
\end{equation}
Clearly, $\tilde{H}$ is ``\cP\cT-symmetric'' under the same metric $W$ as $H$.  We call the second part $A$  ``\cP\cT-anti-symmetric'' with respect to the metric operator $W$. In Appendix A, we show that for any given $\Lambda$ and $W$, this partition is unique.

Using the partition in (\ref{eqn:part}), we rewrite ({\ref{eqn:Wdot}) into
\begin{equation}
\ri \hbar \dot{W} = A^\dag W - W A = -2 W A.
\end{equation}
This directly gives
\begin{equation}
A = - \half\ri \hbar W^{-1} \dot{W}.
\end{equation}
Therefore, for a known $W(t)$,  the ``\cP\cT-anti-symmetric'' part of a general time-displacement generator $\Lambda$ can be explicitly worked out. Summarizing, for a time-dependent \cP\cT-symmetric Hamiltonian $H(t)$ which is diagonalizable with real eigenvalues (e.g., within a certain time window before \cP\cT-symmetry breaking), one may find a time-dependent metric $W(t)$, and the generator of time displacements
\begin{equation}
\Lambda = \tilde{H} - \half\ri \hbar W^{-1} \dot{W}
\end{equation}
that yields unitary evolution, where $\tilde{H}$ is always \cP\cT-symmetric with respect to $W(t)$.

It is important to stress that the above unitary condition does not lead to any conclusion about the relation between $H$ and $\tilde{H}$ (other than that they share the same metric $W$). This is not a surprise. Indeed, in convention QM, {\it any}\ Hermitian time-displacement generator leads to a unitary evolution. Taking the Hamiltonian operator as the generator of time displacements in conventional QM is a postulate, not a result by deduction \cite{Dirac}. With this recognition, we proceed to propose an axiom for time-dependent PTQM, i.e.,
\begin{equation}
 H=\tilde{H},
\end{equation}
which leads to
\begin{equation}
\Lambda  = H-\half\ri \hbar W^{-1} \dot{W}.
\end{equation}
That is, {\it the \cP\cT-symmetric part of the generator of time displacements in PTQM should be the same Hamiltonian operator that gives the (instantaneous) energy spectrum of the same system}. We finally have the following time-dependent Schr\"odinger-like equation for PTQM,
\begin{equation}
\ri \hbar \frac{\rd}{\rd t} |\Psi\> = \left( H - \frac{\ri\hbar}{2}W^{-1} \dot{W} \right)|\Psi\>.
\label{eqn:PT_schr}
\end{equation}
This axiom lifts the ambiguity issue in PTQM, first raised in our previous work \cite{GW}. The resulting generator of time evolution in PTQM also differs from what was studied in \cite{GW}.  In addition, it can be seen that conventional QM can now be regarded as a special case of PTQM with the metric operator chosen to be unity at all times.

\subsection{Mapping between conventional QM and time-dependent PTQM}

Since both conventional QM and our time-dependent PTQM generate unitary evolutions, both yield real instantaneous spectra, one may wonder whether there is a simple mapping between these two frameworks. If there is such a mapping, then why do we still need time-dependent PTQM?  In this subsection we shed light on these questions.

As a metric operator, $W$ is a positive-definite Hermitian operator and hence it can be written as
\begin{equation}
W=\eta^\dag\eta,
\label{eqn:Weta}
\end{equation}
where $\eta$ is invertible. Note that the solution to $\eta$ is not unique and as a matter of fact,
\begin{equation}
W = \eta^\dag\eta = \left(U\eta\right)^\dag\left(U\eta\right)\quad {\rm with}\quad U^{-1} = U^\dag.
\end{equation}
So any known solution multiplied by an arbitrary unitary operator (unitary in the sense of conventional QM) from the left is still a solution to $\eta$. One may now use a known $\eta$ to map a \cP\cT-symmetric operator $H$ to a Hermitian operator $h$ by the following similarity transformation,
\begin{equation}
h\equiv \eta H \eta^{-1},
\end{equation}
with $ h=h^\dag$ being a direct consequence of the \cP\cT\ symmetry of $H$ [see (\ref{eqn:Hermitianh})]. As such $h$ can be seen as a Hamiltonian in conventional QM, always with the same spectrum as $H$ in PTQM. The associated eigenkets can also be mapped accordingly.  Specifically, an eigenket $|\psi_n\>$ of $H$ is mapped to an eigenket of $h$ with the same eigenvalue, i.e., for $|\phi_n\> \equiv \eta |\psi_n\>$,
\begin{equation}
H |\psi_n\> = E_n |\psi_n\> \quad \Rightarrow \quad h |\phi_n\> = E_n |\phi_n\>.
\end{equation}

We next extend this mapping via $\eta$ to general time evolving states as well, i.e.,
\begin{equation}
|\Phi\> \equiv \eta |\Psi\>,
\end{equation}
where $|\Psi\>$ ($|\Phi\>$) represents the time evolving states in time-dependent PTQM (conventional QM). Using our general Schr\"odinger-like equation for PTQM in (\ref{eqn:PT_schr}), one finds the following equation of motion for a mapped state in conventional QM,
\begin{equation}
\ri \hbar \frac{\rd}{\rd t} |\Phi\> = \bar{h} |\Phi\>
\quad {\rm with} \quad
\bar{h} = h + \frac{\ri\hbar}{2}\left[\dot{\eta}\eta^{-1} - \left(\dot{\eta}\eta^{-1}\right)^\dag\right].
\label{eqeta}
\end{equation}
Because $\dot{\eta}\eta^{-1}$ is not Hermitian in general, $ h\ne \bar{h}$ and the above mapped equation of motion is in general different from the standard Schr\"odinger equation in (\ref{eqn:full_schr}) for the mapped Hamiltonian $h$.

Equation (\ref{eqeta}) indicates that among all possible choices for the mapping operator $\eta$, the one with
\begin{equation}
\dot{\eta}_{\rm proper}\eta^{-1}_{\rm proper} = \left(\dot{\eta}_{\rm proper}\eta^{-1}_{\rm proper}\right)^\dag
\label{eqn:proper}
\end{equation}
is special because then, the mapped Hamiltonian operator $h$ can afford to play a dual role as expected from conventional QM: It not only gives the spectrum of the mapped system, but also becomes the generator of time displacements for the mapped time evolving states. To emphasize such a peculiar type of mapping, we call this mapping a ``proper'' one.  In this terminology, our time-dependent PTQM is equivalent to  certain Hermitian problems in conventional QM under a proper mapping.  Note however, it can be challenging to find this proper mapping explicitly, a strong indication that time-dependent PTQM is not a trivial extension of conventional QM.

To better understand the complexity of the proper mapping, we first discuss one possible procedure to find the proper mapping.  Suppose an unknown $\eta_{\rm proper}$ is a proper mapping satisfying \eqn{eqn:proper}. Then another known mapping with
\begin{equation}
\eta = U \eta_{\rm proper}
\label{eqn:improper}
\end{equation}
is found to satisfy
\begin{equation}
\dot{U}=\half \left[\dot{\eta}\eta^{-1} - \left(\dot{\eta}\eta^{-1}\right)^\dag \right] U.
\label{eqn:properU}
\end{equation}
Seeking the proper mapping $\eta_{\rm proper}$ itself is now converted to solving the above first-order differential equation for $U$. This can be easier than directly searching for a Hermitian $\dot{\eta}_{\rm proper}\eta^{-1}_{\rm proper}$. The suggested procedure hence goes like the following.  First, we find a special solution to (\ref{eqn:Weta}), denoted by $\eta$. Then we solve (\ref{eqn:properU}) for $U$. Finally, $\eta_{\rm proper} = U^\dag\eta$ gives us a proper mapping. Because the right hand side of ({\ref{eqn:properU}) is time-dependent, in general there is no analytical solution to $U$.  In addition, there are no boundary conditions for $U$ with respect to time [such as $U|_{t=0}=\one$], so the general solution to (\ref{eqn:properU}) has an arbitrary constant unitary factor. This factor corresponds to the unitary equivalence in the mapped Hermitian system, namely, the arbitrariness of choosing a basis to write down the Hamiltonian.

For a proper mapping $\eta_{\rm proper}$, the Sch\"odinger-like equation in (\ref{eqn:PT_schr}) takes the form
\begin{equation}
\ri \hbar \frac{\rd}{\rd t} |\Psi\> = \left( H - \ri\hbar \eta^{-1}_{\rm proper} \dot{\eta}_{\rm proper} \right)|\Psi\>.
\label{eqn:znojil}
\end{equation}
An equation of this form was also studied in \cite{Znojil} and implicitly in \cite{Fring}. However, we should stress that our perception and approach are distinct from \cite{Fring} and \cite{Znojil}. For example, in \cite{Znojil}, the author started from a Hermitian Hamiltonian $h(t)$, and then applied a time-dependent similarity transformation to obtain $H(t) = \eta^{-1}(t) h(t) \eta(t)$ (Note that $\eta$ was called $\Omega$ there) and the equation of motion (\ref{eqn:znojil}). In such a treatment, a \cP\cT-symmetric system is derived from a Hermitian system (so everything is known from the Hermitian system). On the contrary, we treat a non-Hermitian \cP\cT-symmetric Hamiltonian as a fundamental starting point. Any Hermitian Hamiltonian is a special case of a more general \cP\cT-symmetric Hamiltonian. The conventional time-dependent QM can be derived from time-dependent PTQM. That is, the conventional time-dependent Schr\"odinger equation can be reduced from our general axiom in (\ref{eqn:PT_schr}) when the metric operator is set to be the unit operator. In this view, a proper mapping between a \cP\cT-symmetric system and a Hermitian system is to be sought (which can be challenging), and it is not known {\it a priori}.

Before ending this subsection, we also note that (\ref{eqn:properU}) is equally applicable to a Hermitian system in conventional QM. In that case, $W=\one$, and any unitary operator would be a solution to $\eta$ in (\ref{eqn:Weta}) and the simplest solution $\eta_{\rm proper}=\one$ is a proper mapping. Equation (\ref{eqn:properU}) then indicates that an ``improper'' mapping $\eta$ shifts the Hamiltonian by
\begin{equation}
\half\ri\hbar\left[\dot{\eta}\eta^{-1} - \left(\dot{\eta}\eta^{-1}\right)^\dag\right]= \ri\hbar \dot{U} U^\dag = -\ri\hbar U \dot{U}^\dag,
\end{equation}
which is precisely the expected shift to the time-displacement generator in conventional QM under a time-dependent unitary (gauge) transformation [as already seen in (\ref{eqn:SchrU})].

\section{Example: A complex harmonic oscillator}
\label{sec:CHO}

In this section we wish to use a very simple example to illustrate time-dependent PTQM.  Though in general it is challenging to identify how such time-dependent problems can be properly mapped onto a familiar problem in conventional QM,  we have made efforts to construct examples where such a proper mapping can be found.

Consider then a complex quadratic Hamiltonian
\begin{equation}
H_{\rm CHO} = \frac{1}{2} \left[ \left( X + 2\ri y \frac{Y}{Z}- y^2 \frac{Y^2}{Z}\right)\hat{q}^2 + \left(Y+\ri y\right)(\hat{p}\hat{q}+\hat{q}\hat{p}) + Z \hat{p}^2\right],
\label{eqn:CHO}
\end{equation}
where $X$, $Y$, $Z$, and $y$ are real, possibly time-dependent parameters, with $X$, $Z$, and $y$ positive and $ZX>Y^2$. It is straightforward to find that the eigenvalues of this complex Hamiltonian are all real and positive, which are given by
\begin{equation}
E_n = \left( n+ \half\right) \hbar \sqrt{ZX-Y^2}\quad {\rm with}\quad n=0,1,2,\cdots.
\end{equation}
One also finds $H_{\rm CHO}$ in \eqn{eqn:CHO} being self-adjoint with respect to the following metric operator
\begin{equation}
W=\exp\left( -\frac{1}{\hbar} \frac{y}{Z} \hat{q}^2\right).
\label{Wex}
\end{equation}
According to our axiom proposed above, we then obtain the time-displacement generator $\Lambda$ for this system,
\begin{equation}
\qquad \Lambda =  \frac{1}{2} \left\{ \left[ X + 2\ri y \frac{Y}{Z} - y^2 \frac{Y^2}{Z} + \ri \frac{\rd}{\rd t} \left(\frac{y}{Z}\right)\right] \hat{q}^2  + \left(Y+\ri y\right)(\hat{p}\hat{q}+\hat{q}\hat{p}) + Z^2 \hat{p}^2\right\}.
\end{equation}

To digest the above expression for $\Lambda$, we now attempt to find a proper mapping to identify an equivalent time-dependent problem in conventional QM. To that end we factorize $W$ as $W=\eta_{\rm proper}^\dag\eta_{\rm proper}$. As seen in (\ref{eqn:improper}), any other mapping $\eta$ can be expressed as a product of a unitary operator and the proper mapping, $\eta=U\eta_{\rm{proper}}.$ Using (\ref{Wex}), one can easily find one Hermitian square-root of $W$, i.e.,
\begin{equation}
\eta = \exp\left(-\frac{1}{2\hbar} \frac{y}{Z}\hat{q}^2\right) = \eta^\dag.
\label{eqn:Hermitianeta}
\end{equation}
To find a proper mapping, we then consider the following {\it ansatz}\ for the unitary factor,
\begin{equation}
U = \exp\left[\frac{\ri}{\hbar} \left(\frac{\kappa}{2}\hat{q}^2 + \upsilon\right) \right],
\end{equation}
where $\kappa$ and $\upsilon$ are real coefficients to be determined.

The proper mapping operator $\eta_{\rm{proper}}=U^\dag \eta$ maps $H_{\rm CHO}$ onto a Hermitian Hamiltonian,
\begin{eqnarray}
h_{\rm CHO} &=& \eta_{\rm{proper}}H_{\rm CHO} \eta_{\rm{proper}}^{-1}\nn
&=& \half \left[ \left( X + 2\kappa Y + \kappa^2 Z \right) \hat{q}^2
 + \left(Y+ \kappa Z\right)(\hat{p}\hat{q}+\hat{q}\hat{p}) + Z \hat{p}^2\right].
\end{eqnarray}
The mapped Schr\"odinger equation in conventional QM has the following time-displacement generator,
\begin{eqnarray}
\bar{h}_{\rm CHO} &=& h_{\rm CHO} + \half \ri \hbar\left[ \dot{\eta}\eta^{-1} - \left(\dot{\eta}\eta^{-1}\right)^\dag \right]\nn
&=& h_{\rm CHO} + \half \dot{\kappa}\hat{q}^2 +\dot{\upsilon}.
\end{eqnarray}
Since $\eta_{\rm{proper}}$ is a proper mapping, we must have $h_{\rm CHO} = \bar{h}_{\rm CHO}$. So $\kappa$ and $\upsilon$ need to be chosen as constants in time. As a result, $U$ is a constant unitary operator. For convenience we may choose $\kappa=0$ and $\upsilon=0$. In other words, the Hermitian mapping operator $\eta$ in (\ref{eqn:Hermitianeta}) is actually a proper one. Finally, the time evolution of a complex harmonic oscillator is properly mapped to that of the so-called generalized harmonic oscillator in conventional QM, namely,
\begin{equation}
h_{\rm GHO} = \half \left[ X \hat{q}^2 + Y(\hat{p}\hat{q}+\hat{q}\hat{p}) + Z \hat{p}^2\right].
\end{equation}
A similar result on the special case of $Y=0$ was first obtained in \cite{Fring}. It should be stressed that this example is specially designed to demonstrate the consistency of time-dependent PTQM. In general it is highly demanding to find the analytic form of a proper mapping.

\section{Berry phase in time-dependent PTQM: How a continuously tunable fictitious magnetic monopole emerges}
\label{sec:Berry}

In this section we focus on adiabatic evolution and the associated Berry phase in a class of \cP\cT-symmetric two-level systems. When considering adiabatic manipulation, the Hamiltonian changes with time and so does the metric operator in general.  As such, this kind of setup becomes a test bed to apply our time-dependent PTQM.  Since Berry-phase problem in conventional QM is well known, it will be also fruitful to make a comparison between predictions from time-dependent PTQM and the familiar results in conventional QM. In particular, we shall re-examine the surprising findings made in \cite{GW} (such as a fictitious nontrivial Dirac string and a fictitious continuously tunable monopole) using our new formalism.  By a proper mapping from a \cP\cT-symmetric system to a conventional quantum mechanical system, we are able to clearly explain the origin of our previous findings.   Thus the main purpose of this section is not to carry out similar calculations as in \cite{GW} (such calculations are nevertheless necessary because the time-displacement generator here differs from that adopted in \cite{GW}), but to provide important insights to understand why a fictitious nontrivial Dirac string and a fictitious continuously tunable monopole become possible in time-dependent PTQM.

Let us first consider a general time-dependent PTQM problem with the Schr\"odinger-like equation in (\ref{eqn:PT_schr}), where $H$ is a time-dependent \cP\cT-symmetric Hamiltonian that determines the instantaneous spectrum of the system. The explicit time dependence of $H$ is implemented via a set of time evolving parameters ${\bf X}\equiv (X_1, X_2, X_3, \cdots)$ such that $H=H[{\bf X}(t)]$.  The associated $W$ can then be understood as $W=W[{\bf X}(t)]$.

We expand the solution of (\ref{eqn:PT_schr}), $|\Psi(t)\>$ in terms of the instantaneous eigenstates of $H[{\bf X}(t)]$,
\begin{equation}
|\Psi(t)\> = \sum_n c_n(t) \e^{\ri\alpha_n(t)} |\psi_n(t)\>,
\label{eqn:Psipsi}
\end{equation}
where $\alpha_n$ is a dynamical phase defined as
$\alpha_n(t) = -\frac{1}{\hbar}\int^t \rd \tau\, E_n[{\bf X}(\tau)]$.
If we make an adiabatic approximation by ignoring the population transitions, we have
\begin{equation}
c_n(t) \approx c_n(0) \e^{\ri\gamma_n^{g}(t)},
\end{equation}
where $\gamma_n^{g}$ is identified as the geometric phase and it is given by
\begin{equation}
\gamma_n^{g} = \ri  \int\rd {\bf X}\cdot \left[ \<\psi_n|W \nabla|\psi_n\> + \half\<\psi_n|(\nabla W)|\psi_n\> \right],
\label{gephase1}
\end{equation}
where $\nabla \equiv \frac{\p}{\p{\bf X}}$. A very similar result was first obtained by the same authors in \cite{GW}. But the derivation there is rather brief. For the sake of completeness, a detailed derivation using our proposed equation of motion (\ref{eqn:PT_schr}) is given in \ref{sec:PTQM}.

The geometric phase associated with a closed adiabatic path yields a Berry phase in our time-dependent PTQM. Interestingly, the first term in (\ref{gephase1}) represents an integral over a Berry connection (as expected, the connection as an inner product carries a metric $W$).  On top of that, there is a second term, which originates from the metric's dependence on system parameters. Only the combination of the two terms results in a real phase $\gamma_n^g$.

\subsection{From Berry phase results to a continuously tunable magnetic monopole}

Now let us apply these general results to a $2\times2$ \cP\cT-symmetric system with the Hamiltonian \cite{WCZ},
\begin{equation}
H_{2\times2} = \epsilon\one_{2\times2} + \left( a\, {\bf n}^r + \ri b\sin\delta\, {\bf n}^\theta + \ri b\cos\delta\, {\bf n}^\varphi \right)\cdot{\bm \sigma},
\label{Hform}
\end{equation}
where all six parameters, $\epsilon$, $a$, $b$, $\theta$, $\varphi$, and $\delta$ are real, ${\bm \sigma}$ are Pauli matrices and three mutually orthogonal unit vectors are defined as
\begin{eqnarray}
{\bf n}^r &\equiv& (\sin\theta\cos\varphi,\sin\theta\sin\varphi,\cos\theta), \nn
{\bf n}^\theta &\equiv& (\cos\theta\cos\varphi,\cos\theta\sin\varphi,-\sin\theta), \nn
{\bf n}^\varphi &\equiv& (-\sin\varphi,\cos\varphi,0).
\end{eqnarray}
Obviously, $H_{2\times2}$ is Hermitian if and only if $b=0$.

The eigenvalues of the above Hamiltonian, $E_\pm = \epsilon \pm \sqrt{a^2 - b^2}$ are real when $a^2 \geq b^2$. For our later calculations, we limit our discussions in the regime of $a^2 > b^2$ where $H_{2\times2}$ is always diagonalizable with two non-degenerate real eigenvalues (\cP\cT-symmetry is not spontaneously broken). The associated metric operator can be chosen as
\begin{equation}
W = \one_{2\times2} + \frac{b}{a}\left(\cos\delta\, {\bf n}^\theta - \sin\delta\, {\bf n}^\varphi \right)\cdot{\bm \sigma}.
\label{eqn:simpleW}
\end{equation}

The expressions for two geometric phases $\gamma_{\pm}^{g}$ associated with two adiabatic eigenstates are surprisingly simple (please see \ref{sec:2x2Berry} for detailed derivations)
\begin{equation}
\gamma_{\pm}^{g}=  \int \left( F_\pm^{\varphi}\ \rd\varphi +   F_{\pm}^{\theta}\ \rd\theta +   F_{\pm}^\delta\ \rd\delta\right),
\label{gammaeq2}
\end{equation}
with
\begin{eqnarray}
F_{\pm}^{\varphi} &=& \frac{1}{2} \left( 1 \pm \frac{a}{\sqrt{a^2-b^2}}\cos\theta\right),\nn
F_{\pm}^{\theta} &=&  \frac{1}{2}, \nn
F_{\pm}^{\delta} &=& \pm \frac{1}{2} \left( 1 - \frac{a}{\sqrt{a^2-b^2}}\right).
\label{ftheta}
\end{eqnarray}

For Berry phase generation via the adiabatic parameter pair $(\theta, \varphi)$, we adopt a language similar to our previous work \cite{GW}. To proceed we assume that $(\theta, \varphi)$ are the two angles in a standard spherical coordinate system, $(r,\theta,\varphi)$.  We denote $\gamma_\pm^B$ as the Berry phase generated after the system has adiabatically moved along a closed path on a sphere,
\begin{equation}
\gamma_\pm^B=  \oint \left( F_\pm^\varphi \rd\varphi +   F_\pm^\theta \rd\theta \right)=\frac{e}{\hbar} \iint {\bf B}_\pm\cdot \rd{\bf S},
\label{eqn:Berryphase1}
\end{equation}
where the fictitious magnetic field is given by
\begin{equation}
{\bf B}_\pm = \frac{\pi\hbar}{e}\left(1 \pm \frac{a}{\sqrt{a^2-b^2}}\right)\delta(x)\delta(y) {\bf n}^z \mp \frac{\hbar}{2e} \frac{a}{\sqrt{a^2-b^2}} \frac{\bf r}{r^3},
\label{eqn:fictB1}
\end{equation}
with ${\bf n}^z\equiv{\bf n}^r|_{\theta=0}$. This virtual magnetic field describes a monopole with charge $\left(\mp\frac{a}{\sqrt{a^2-b^2}}Q_m\right)$ at the origin and a string component along the $\theta=0$  ($z$-) direction, where $Q_m\equiv \frac{2\pi\hbar}{\mu_0 e}$ is the conventional magnetic monopole charge \cite{monopole}.  In sharp contrast to the well-known Berry phase problem of a two-level system in conventional QM, here the fictitious magnetic monopole is no longer quantized: its charge is continuously tunable between $(-\infty,-Q_m]$ and $[Q_m,\infty)$.  Echoing with this, the phase contribution from the string component is no longer an integer multiple of $2\pi$ and hence becomes an observable quantity. This is again drastically different from the trivial Dirac string in conventional QM, which can only produce an unobservable phase.

\subsection{Proper mapping to Hermitian two-level systems}

The Berry phase results in the previous subsection are surprising and call for more analysis and theoretical insights. Consider then a proper mapping from the above \cP\cT-symmetric two-level problem to a Hermitian one in conventional QM. We proceed by first choosing a Hermitian similarity transformation operator $\eta$, the square-root of $W$. For $W$ in (\ref{eqn:simpleW}), it has two distinct square-roots (on top of the overall $\pm$ signs) \cite{WCZ},
\begin{equation}
 \qquad \eta^{(\pm)} = \frac{1}{\sqrt{2a\chi^{(\pm)}}}\left(
\begin{array}{cc}
\chi^{(\pm)} - b \sin\theta\cos\delta & b (\cos\theta\cos\delta + \ri \sin\delta) \e^{-\ri\varphi}\\
b (\cos\theta\cos\delta - \ri \sin\delta) \e^{\ri\varphi} & \chi^{(\pm)} + b \sin\theta\cos\delta
\end{array}
\right)
\label{eqn:eta}
\end{equation}
with $\chi^{(\pm)}\equiv a\pm\sqrt{a^2-b^2}$. To avoid confusion, we always use {\it subscripts}\ ${}_\pm$ to label eigenvalues $E_\pm$ and use {\it superscripts}\ ${}^{(\pm)}$ to label the different choices in $\eta^{(\pm)}$, the square-roots of $W$.

Neither $\eta^{(+)}$ nor $\eta^{(-)}$ is a proper mapping as defined in (\ref{eqn:proper}). That is, the mapped equation of motion will not be governed by $h^{(\pm)}\equiv \eta^{(\pm)} H_{2\times2} \left[\eta^{(\pm)}\right]^{-1}$. To find a proper mapping, we need to solve (\ref{eqn:properU}) for the unitary matrix $U$ which links a proper mapping $\eta_{\rm proper}$ to the current known (improper) mapping $\eta^{(\pm)}$ by (\ref{eqn:improper}).

For an arbitrary path with varying $\theta$, $\varphi$, and $\delta$, analytically obtaining such $U$ to establish a proper mapping is impossible. Here, we consider paths that involves changes in $\varphi$ only. Paths with varying just one of the other two angular variables, $\delta$ and $\theta$ can be found in \ref{sec:mapping}. When only $\varphi$ is changing in time, (\ref{eqn:properU}) gives
\begin{eqnarray}
 \frac{\rd U(\varphi)}{\rd\varphi}
 &=& \ri \zeta^{(\pm)} \left(
\begin{array}{cc}
\cos^2\theta\cos^2\delta + \sin^2\delta &
\sin\theta\cos\delta(\cos\theta\cos\delta + \ri\sin\delta )\e^{-\ri\varphi}\\
\sin\theta\cos\delta(\cos\theta\cos\delta + \ri\sin\delta )\e^{\ri\varphi} &
-\cos^2\theta\cos^2\delta - \sin^2\delta
\end{array}
\right)U(\varphi)\nn
&=& \ri\zeta^{(\pm)} \e^{-\ri\frac{\varphi}{2}\sigma_3} \e^{-\ri\frac{\theta}{2}\sigma_2} \e^{\ri\frac{\delta}{2}\sigma_3}
\left(
\begin{array}{cc}
-\cos\theta & -\ri\sin\theta\sin\delta\\
\ri \sin\theta\sin\delta & \cos\theta
\end{array}
\right)
\e^{-\ri\frac{\delta}{2}\sigma_3} \e^{\ri\frac{\theta}{2}\sigma_2} \e^{\ri\frac{\varphi}{2}\sigma_3} U(\varphi),
\label{eqn:dUphi}
\end{eqnarray}
where $\zeta^{(\pm)} \equiv \frac{1}{2} \left(1\mp\frac{a}{\sqrt{a^2-b^2}}\right)$. The ranges of $\zeta^{(\pm)}$ are $\zeta^{(+)}\leq 0$ and $\zeta^{(-)}\geq 1$, where the equal signs are only taken at the Hermitian limit, $b\to0$. Use a technique discussed in \ref{sec:mapping}, we obtain the solution for $U(\varphi)$,
\begin{equation}
 U(\varphi)
= \e^{-\ri\frac{\varphi}{2}\sigma_3}
  \exp\left\{\ri\left[\zeta^{(\pm)} \e^{-\ri\frac{\theta}{2}\sigma_2} \e^{\ri\frac{\delta}{2}\sigma_3}
   \left(
    \begin{array}{cc}
    -\cos\theta & -\ri \sin\theta\sin\delta\\
    \ri \sin\theta\sin\delta & \cos\theta
    \end{array}
  \right)\e^{-\ri\frac{\delta}{2}\sigma_3} \e^{\ri\frac{\theta}{2}\sigma_2}+ \frac{1}{2}\sigma_3 \right]\varphi\right\} U_0.
\end{equation}
For generic values of $\theta$ and $\delta$, the mapped Hermitian Hamiltonian is too complicated to be written down here. Instead, we only give the explicit forms for the special cases of $\delta=0$ and $\delta=\pi/2$ and choose to numerically plot a generic mapping in Figure \ref{fig:Phiphi} and Figure \ref{fig:map}. For simplicity, we choose $U_0=\one_{2\times2}$ in the mapping.
\begin{itemize}
  \item $\delta=\pi/2$.
    In this case, the mapped Hamiltonian is very simple:
    \begin{equation}
    h^{(\pm)}(\varphi,\delta=\frac{\pi}{2})=\left(
        \begin{array}{cc}
        \epsilon & 0\\
        0 & \epsilon
        \end{array}
    \right)\pm\sqrt{a^2-b^2}\left(
        \begin{array}{cc}
        \cos\theta & \sin\theta\, \e^{-\ri\left[1-2\zeta^{(\pm)}\right]\varphi}\\
        \sin\theta\, \e^{\ri\left[1-2\zeta^{(\pm)}\right]\varphi} & -\cos\theta
        \end{array}
    \right).
    \end{equation}
    It is clear how the proper mapping changes the periodicity of the system. For a non-integer $2\zeta^{(\pm)}$, the mapped Hamiltonian does not come back to its original value when $\varphi$ changes from $0$ to $2\pi$. Since $\zeta^{(+)}\leq 0$ and $\zeta^{(-)}\geq 1$, the factor $\left|1-2\zeta^{(\pm)}\right|$ is always greater or equal to one. This means that the period of mapped Hamiltonian, $2\pi/\left|1-2\zeta^{(\pm)}\right|$ is always less than or equal to that of the \cP\cT-symmetric system.
  \item $\delta=0$.
    We parameterize the mapped Hermitian Hamiltonian as
    \begin{equation}
    h^{(\pm)}(\varphi,\delta=0)=\left(
        \begin{array}{cc}
        \epsilon & 0\\
        0 & \epsilon
        \end{array}
    \right) \pm \sqrt{a^2-b^2}\left(
    \begin{array}{cc}
    \cos[\Theta(\varphi)] & \sin[\Theta(\varphi)]~ \e^{-\ri\Phi(\varphi)}\\
    \sin[\Theta(\varphi)]~ \e^{\ri\Phi(\varphi)} & -\cos[\Theta(\varphi)]
    \end{array}
    \right).
    \label{eqn:hmapped}
    \end{equation}
    In this case, the angular parameters in the mapped Hamiltonian are
    \begin{eqnarray}
    \cos[\Theta(\varphi)] &=& \cos\theta \left\{1 - \frac{2\zeta^{(\pm)}}{\xi^2} \sin^2\theta [1-\cos (\xi\varphi)]\right\},\nn
    \sin[\Theta(\varphi)]\,\e^{-\ri\Phi(\varphi)} &=& \sin\theta \left\{1 - \frac{1}{\xi^2} \left[1 - 2 \zeta^{(\pm)}\cos^2\theta\right] [1-\cos \left(\xi\varphi\right)] -\frac{\ri}{\xi} \sin\left(\xi\varphi\right)\right\},
    \label{eqn:mappedhphi}
    \end{eqnarray}
    where $\xi\equiv\sqrt{1+4\zeta^{(\pm)}\left(\zeta^{(\pm)}-1\right)\cos^2\theta}$. The same key feature of the mapped Hamiltonian appears: Though it remains periodic functions of $\varphi$, it has a period different from $2\pi$. Since $\xi\ge1$, the mapped Hamiltonian always has a shorter period, $2\pi/\xi$.
  \item Generic $\delta$.
    The expressions of the mapped Hamiltonian is too complicated to be included here. Therefore, we solve the angular parameters $\Theta$ and $\Phi$ numerically for a special case with $\zeta^{(+)}=-0.95$, $\theta=\pi/3$, and $\delta=10\pi/9$ and plot the results in two figures. In Figure \ref{fig:Phiphi}, we plot $\Phi$ as a function of $\varphi$. It clearly shows that $\Phi$ increases faster than $\varphi$. In particular, $\Phi$ changes from $0$ to $2\pi$ as $\varphi$ changes from $0$ to $1.07\pi$. In Figure \ref{fig:map}, we plot two mapped paths on the same unit sphere. The $(\theta,\varphi)$ path is plotted with
    the dashed red line and the $(\Theta,\Phi)$ path is plotted with the solid blue line. The $(\theta,\varphi)$ path is horizontal because $\theta$ is fixed; whereas the $(\Theta,\Phi)$ path is not horizonal in general as both $\Theta$ and $\Phi$ change with $\varphi$.  As the $(\Theta,\Phi)$ path completes a circle on the sphere, we see that the parameter $\varphi$ only changes from $0$ to $1.07\pi$, thus forming an open path instead. The proper mapping between an adiabatic process in PTQM and an adiabatic process in conventional QM is thus subtle.

    \begin{center}
    \begin{figure}[htb]
    \caption{Angular parameter $\Phi$ in the mapped Hamiltonian as a function of the angular parameter $\varphi$ in the \cP\cT-symmetric Hamiltonian $H_{2\times2}$. Here we choose parameter $\zeta^{(+)}=-0.95$ and fix the two other angular parameters at $\theta=\pi/3$ and $\delta=10\pi/9$. This plot shows that $\Phi$ grows faster than $\varphi$ and $\Phi(\varphi\approx 1.07\pi)=2\pi$.     }
    \includegraphics[width=150mm]{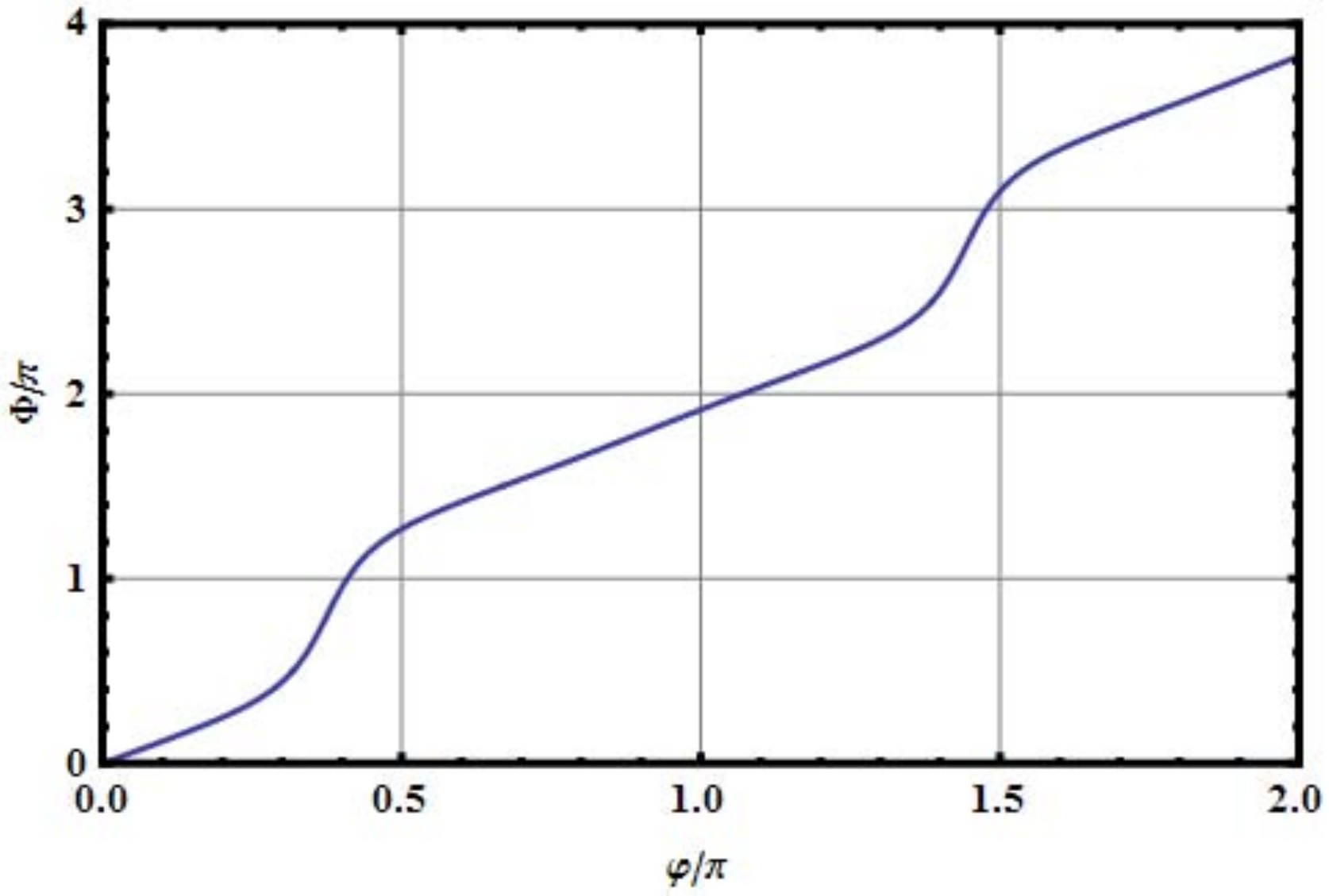}
    \label{fig:Phiphi}
    \end{figure}
    \end{center}

    \begin{center}
    \begin{figure}[htb]
    \caption{The adiabatic path in PTQM and the mapped adiabatic path in conventional QM,
     plotted on the same unit sphere. The red dashed line is an open path in $(\theta,\varphi)$ parameter space of the \cP\cT-symmetric Hamiltonian $H_{2\times2}$ with $\theta=\pi/3$ and $\varphi$ varies from $0$ to $1.07\pi$. The blue solid line is the mapped closed path in $(\Theta,\Phi)$ parameter space of the mapped Hamiltonian. In the plot we choose parameters $\zeta^{(+)}=-0.95$ and $\delta=10\pi/9$.}
    \includegraphics[width=150mm]{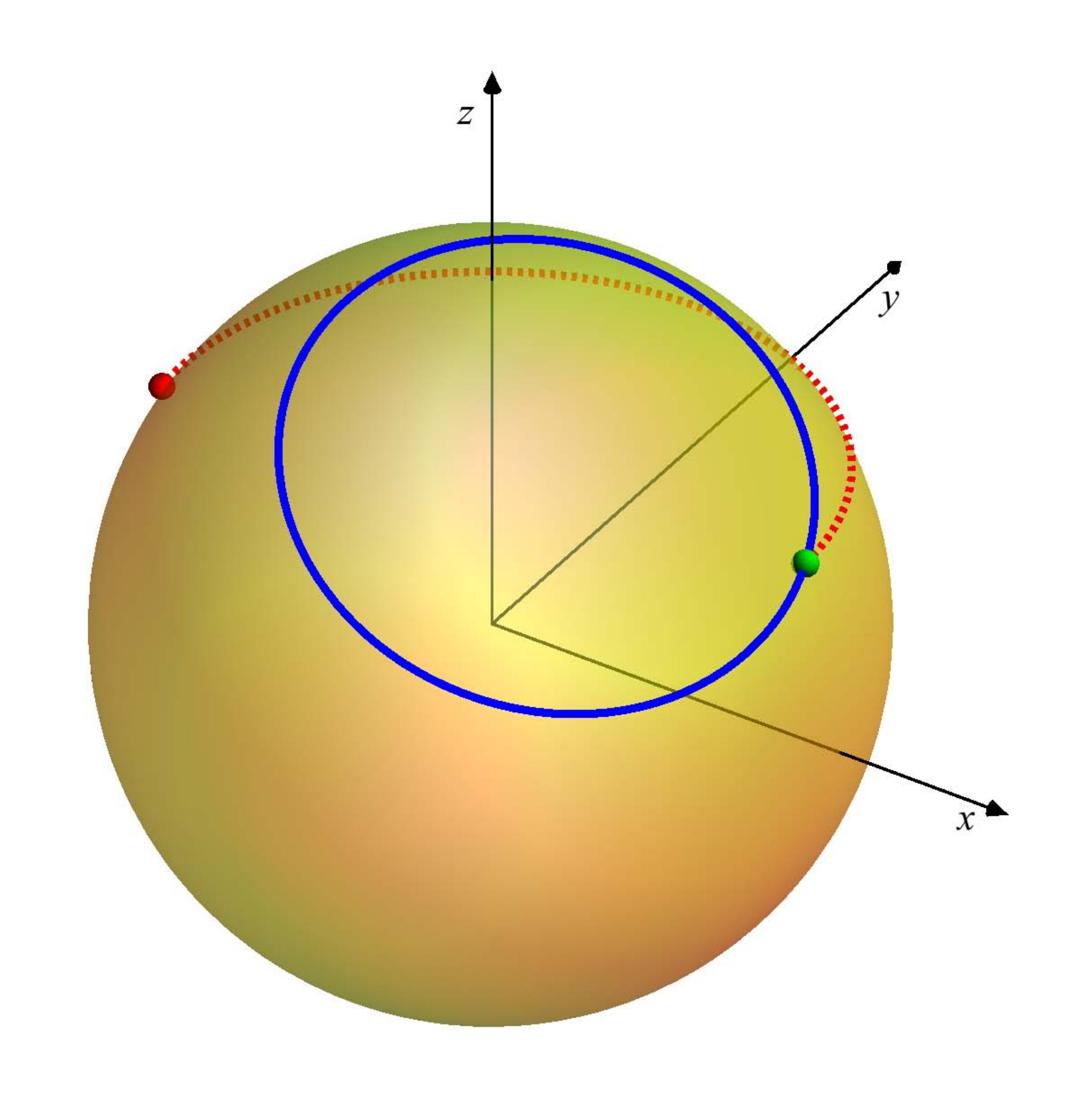}
    \label{fig:map}
    \end{figure}
    \end{center}
   \end{itemize}

In summary, it is clear that the properly mapped Hamiltonian in general is {\it not}\ a period-$2\pi$ function of the angular parameters $\theta$, $\varphi$, or $\delta$.  From this viewpoint, our previous Berry phase results can be better understood. The Berry phase for $H_{2\times2}$ in our PTQM is the geometric phase associated with a {\it closed}\ path in the $(\theta,\varphi,\delta)$ parameter space, which is in general an {\it open}\ path in the properly mapped $(\Theta,\Phi)$ space.  Hence, in the language of a properly mapped Hermitian two-level Hamiltonian [e.g., $h^{(\pm)}(\varphi)$ in (\ref{eqn:hmapped})], our Berry phase results in PTQM problems can at most be translated into a geometric phase associated with an {\it open}\ path in the $(\Theta,\Phi)$ parameter space. Equivalently, a closed path in conventional QM is mapped to an open path in PTQM, \eg, Figure \ref{fig:map}. Therefore, there is nothing inconsistent between our axiom for PTQM and conventional QM. It is also clearer by now, that despite the existence of a proper mapping from time-dependent PTQM to conventional QM, time-dependent PTQM still has its own rich features.

\section{Conclusions}
\label{sec:conclusions}

In this work we treat PTQM as a fundamental theory and consider the conventional QM as a special case of PTQM. We advocate a general framework for time-dependent PTQM and discuss in depth its relation with time-dependent problems in conventional QM. Our study is complementary to current studies of spectral problems in PTQM. The recent experimental interests in PTQM \cite{exp1,exp2,exp3,NaturePhys1,NaturePhys2,Science,Nature} are to mainly simulate PTQM using electromagnetic (optical or microwave) systems. We think that it is possible to simulate or synthesize the Schr\"odinger-like equation proposed in this work and then examine the time evolution, which is unitary if viewed with an appropriate time-dependent metric operator. In particular, the confirmation of our Berry phase results, which may be interpreted as the flux of a continuously tunable fictitious magnetic monopole plus a singular string component, would be of great interest. A much harder and more important experiment needs to be done is to work with a non-Hermitian quantum system directly. We can imagine that experiments may verify our axiom for time-dependent PTQM by (i) directly working on certain non-Hermitian quantum systems with a time-dependent instantaneous real spectrum and a time-dependent metric, and (ii) checking if our proposed Schr\"odinger-like equation describes the actual quantum dynamics. Such experiments would constitute a truly fundamental test of time-dependent PTQM.

\section*{Acknowledgement}
We are especially grateful to Prof.~Adreas Fring for bringing \cite{Fring} to our attention. QW would like to thank Dr.~Daniel Hook for helps on making plots.

\appendix
\section{The uniqueness of the \cP\cT-symmetric and \cP\cT-anti-symmetric partition}
If an operator $H$ is \cP\cT-symmetric with respect to $W=\eta^\dag\eta$, then $\eta H \eta^{-1}$ must be Hermitian:
\begin{equation}
W H = H^\dag W, \quad \Rightarrow\quad \eta H \eta^{-1} = (\eta H \eta^{-1})^\dag.
\label{eqn:Hermitianh}
\end{equation}
Similarly, for a \cP\cT-anti-symmetric operator $A$, $\eta A \eta^{-1}$ is anti-Hermitian:
\begin{equation}
W A = -A^\dag W, \quad \Rightarrow\quad \eta A \eta^{-1} = -(\eta A \eta^{-1})^\dag.
\end{equation}
Now, suppose that $\Lambda$ has two different partitions,
\begin{equation}
\Lambda = \tilde{H} + A = \tilde{H}' + A'.
\end{equation}
Let us sandwich the above equation between $\eta$ and $\eta^{-1}$, obtaining
\begin{equation}
\eta \tilde{H} \eta^{-1} + \eta A\eta^{-1} = \eta\tilde{H}'\eta^{-1} + \eta A'\eta^{-1}.
\end{equation}
Note that both $\eta \tilde{H} \eta^{-1}$ and $\eta \tilde{H}' \eta^{-1}$ are Hermitian and both $\eta A \eta^{-1}$ and $\eta A' \eta^{-1}$ are anti-Hermitian. Because the way of partitioning a matrix into a Hermitian part and an anti-Hermitian part is unique, we must have
\begin{eqnarray}
\eta \tilde{H} \eta^{-1} &=& \eta \tilde{H}' \eta^{-1},\nn
\eta A \eta^{-1} &=& \eta A' \eta^{-1}.
\end{eqnarray}
This means
\begin{equation}
\tilde{H}  =  \tilde{H}' ,\qquad {\rm and} \qquad
 A = A'.
\end{equation}
That is, the way of partitioning $\Lambda$ into a \cP\cT-symmetric part and a \cP\cT-anti-symmetric part is unique for a given $W$.

\section{Adiabatic evolution and the geometric phase in PTQM}
\label{sec:PTQM}

To investigate the adiabatic evolution, let us first take a time derivative on the eigenvalue equation (\ref{eqn:PTeigen}),
\begin{equation}
\dot{H}|\psi_n\> + H|\dot{\psi}_n\> = \dot{E}_n|\psi_n\> + E_n|\dot{\psi}_n\>.
\end{equation}
Multiplying the above equation by $\<\psi_m|W$ from the left and then applying the orthonormal condition in (\ref{eqn:orthonormal}), we arrive at
\begin{equation}
\<\psi_m|W\dot{H}|\psi_n\> = \delta_{mn} \dot{E}_n + (E_n-E_m)\<\psi_m|W|\dot{\psi}_n\>.
\end{equation}
Assuming no degeneracy in our Hamiltonian, the above equation leads to
\begin{equation}
\<\psi_m|W|\dot{\psi}_n\> = \frac{\<\psi_m|W\dot{H}|\psi_n\>}{E_n-E_m},\quad {\rm for} \quad m\ne n,
\label{eqn:mneqn}
\end{equation}
and
\begin{equation}
\dot{E}_n = \<\psi_n|W\dot{H}|\psi_n\>.
\label{eqn:meqn}
\end{equation}
Next, we expand $|\Psi\>$ in terms of the eigenstates of $H$ as in (\ref{eqn:Psipsi}). Taking the time derivative of this expansion we obtain
\begin{equation}
|\dot{\Psi}\> = \sum_n \left( \dot{c}_n - \frac{\ri}{\hbar} c_n E_n \right) \e^{\ri \alpha_n} |\psi_n\> + \sum_n c_n \e^{\ri \alpha_n} |\dot{\psi}_n\>.
\end{equation}
Plugging the above expression into the Shr\"odinger-like equation in (\ref{eqn:PT_schr}) and applying the eigenvalue equation in (\ref{eqn:PTeigen}), we get
\begin{equation}
- \frac{1}{2} W^{-1}\dot{W}\sum_n c_n \e^{\ri \alpha_n} |\psi_n\>= \sum_n \dot{c}_n \e^{\ri \alpha_n} |\psi_n\> + \sum_n c_n \e^{\ri \alpha_n} |\dot{\psi}_n\>.
\end{equation}
 Further multiplying the above equation by $\<\psi_m|W$ from the left and then using the relations in (\ref{eqn:mneqn}) and (\ref{eqn:meqn}), one finds the time dependence of the expansion coefficients $c_m$, namely,
\begin{eqnarray}
\dot{c}_m &=& -c_m \left(\<\psi_m|W|\dot{\psi}_m\> +\frac{1}{2}\<\psi_m|\dot{W}|\psi_m\> \right)\nn
&&\quad + \sum_{n\neq m} c_n\e^{\ri(\alpha_n-\alpha_m)} \left(\frac{\<\psi_m|W\dot{H}|\psi_n\>}{E_n-E_m} + \frac{1}{2}\<\psi_m|\dot{W}|\psi_n\>\right).
\label{eqn:cmdot}
\end{eqnarray}

As in conventional QM, we now make an adiabatic approximation by ignoring the population transition terms (i.e., with $n\ne m$) in the above equation.  This then brings us to
\begin{equation}
c_m(t) \approx c_m(0) \e^{\ri\gamma_m^{g}(t)},
\end{equation}
where the geometric phase $\gamma_m^{g}$ is given by
\begin{equation}
\dot{\gamma}_m^{g}= \ri \left(\<\psi_m|W|\dot{\psi}_m\> +\half\<\psi_m|\dot{W}|\psi_m\> \right).
\label{gephase0}
\end{equation}
In terms of time evolving system parameters ${\bf X}$, the geometric phase is then given by
\begin{equation}
\gamma_m^{g} = \ri  \int\rd {\bf X}\cdot \left[ \<\psi_m|W \nabla|\psi_m\> + \half\<\psi_m|(\nabla W)|\psi_m\> \right],
\label{gephase}
\end{equation}
which is identical to (\ref{gephase1}).

\section{Berry phase in \cP\cT-symmetric two-level systems}
\label{sec:2x2Berry}

For each angular parameter in $H_{2\times2}$, $\theta$, $\varphi$, or $\delta$ , $H_{2\times2}$ is a periodic function with the same period $2\pi$. Since $a\to -a$ is equivalent to $\theta \to \theta \pm \pi$ and $\delta \to 2\pi - \delta$, we may work with $a\geq 0$ without loss of generality. This choice help us to simplify the expressions of eigenstates:
\begin{eqnarray}
|\psi_+\> &=& n_+ \left(
\begin{array}{c}
\left(\cos\frac{\theta}{2} + \beta \e^{-\ri\delta} \sin\frac{\theta}{2}  \right)\e^{-\ri\varphi}\\
-\beta \e^{-\ri\delta} \cos\frac{\theta}{2}  + \sin\frac{\theta}{2}
\end{array}\right),\nn
|\psi_-\> &=& n_- \left(
\begin{array}{c}
-\left(\beta \e^{\ri\delta} \cos\frac{\theta}{2} + \sin\frac{\theta}{2} \right) \e^{-\ri\varphi}\\
\cos\frac{\theta}{2} - \beta \e^{\ri\delta} \sin\frac{\theta}{2}
\end{array}\right),
\label{eqn:eigenstates}
\end{eqnarray}
where $\beta\equiv \frac{b}{a+\sqrt{a^2-b^2}}$. In the Hermitian limit, $\beta\to0$, the above eigenstates reduces to the conventional choice for the Hermitian $2\times2$ Hamiltonian [see (\ref{eqn:Hermitiankets})].

The metric operator $W$ is found to take the following general form,
\begin{equation}
W= \mu \left[ a \one_{2\times2} + \left( \nu\, {\bf n}^r + b\cos\delta\, {\bf n}^\theta - b\sin\delta\, {\bf n}^\varphi \right)\cdot{\bm \sigma} \right],
\end{equation}
where $\mu$ is positive and $\nu$ is sufficiently small, $|\nu|<\sqrt{a^2-b^2}$. The eigenstates in (\ref{eqn:eigenstates}) are orthogonal with respect to this metric operator,
\begin{equation}
\<\psi_+|W|\psi_-\> = 0 = \<\psi_-|W|\psi_+\>.
\end{equation}
The norms of the eigenstates are
\begin{equation}
{\cal N}_\pm \equiv \<\psi_\pm|W|\psi_\pm\> = 2\mu \left|n_\pm\right|^2 \sqrt{a^2- b^2} \frac{\sqrt{a^2-b^2} \pm \nu}{\sqrt{a^2- b^2}+a} .
\end{equation}
Note that ${\cal N}_\pm$ are always positive. That is, the eigenstates are always normalizable in the above ranges of parameters. To normalize the eigenstates, one may either adjust the coefficients $n_\pm$ in $|\psi_\pm\>$, or tune parameters $\mu$ and $\nu$ in $W$. These two ways are equivalent. This means that the undetermined parameters $\mu$ and $\nu$ can always be absorbed by the normalizations of the eigenstates. We choose $\mu=1/a$ and $\nu=0$ for our considerations below. By fixing $\mu$ and $\nu$ in this manner, the parameters in $W$ are the same ones in $H$, \ie, $W=W[{\bf X}(t)]$. With these simplifications, $W$ assumes the form in (\ref{eqn:simpleW}). In the Hermitian limit, $b\to 0$, $W$ reduces to the identity matrix which is the conventional choice. To normalize the eigenstates using the metric in (\ref{eqn:simpleW}), the coefficients in (\ref{eqn:eigenstates}) can be chosen as
\begin{equation}
n_\pm = \e^{-\ri\frac{\theta}{2}} \sqrt{\frac{a^2 + a\sqrt{a^2-b^2}}{2\left(a^2-b^2\right)}},
\end{equation}
where we have explicitly chosen one realization of the rather arbitrary overall phase factor  to ensure that the eigenstates are periodic function of $\theta$ with a period $2\pi$ instead of $4\pi$.

Taking note of both expressions of $H_{2\times2}$ in (\ref{Hform}) and $W$ in (\ref{eqn:simpleW}), we shall consider adiabatic processes in which the three angular parameters $(\theta, \varphi, \delta)$ change with time and an adiabatic cycle is formed in the end.

Applying the general formula in (\ref{gephase}), lengthy calculations eventually lead to the results in (\ref{gammaeq2}) and (\ref{ftheta}). Note that the expression of $F_{\pm}^{\theta}$ in (\ref{ftheta}) depends on the phase factor in the normalization constants $n_\pm$. For example, if we choose the factor to be $\e^{\ri\frac{\theta}{2}}$ in $n_\pm$, then we would have $F_{\pm}^{\theta} = -\half$. However, the difference is not detectable for $\theta$ varying from $0$ to 0 or to $2\pi$. It is also interesting to note that in the case of only $\delta$ varying in time, the second line in (\ref{eqn:cmdot}), the sum over $n\neq m$ vanishes. This means that the result of $F_\pm^\delta$ in (\ref{ftheta}) is independent of the adiabatic limit.

One may notice that the results of $F_+^\varphi$ and $F_+^\theta$ in (\ref{ftheta}) differ from the results obtained in \cite{GW} [(23) and (24)]. There are two reasons. The first reason is in the evolution equation. Here we follow the axiom in (\ref{eqn:PT_schr}). The choice in \cite{GW} is rather arbitrary. Using the current language, the choice made in \cite{GW} accepts $\tilde{H} \neq H$, which is no longer advocated here. The second reason is in the phase factors of the eigenstates. Our choice in (\ref{eqn:eigenstates}) guarantees that both eigenstates are periodic function of $\theta$ with period $2\pi$. By contrast, in \cite{GW}, the parameter $\theta$ was always treated as one of the spherical coordinates varying from $0$ to $\pi$, so the eigenstates in \cite{GW} (see (19) therein) need not to be, and were not chosen to be, periodic function of $\theta$ with period $2\pi$.

Note that $\frac{\p F_\pm^{\theta}}{\p\delta} = \frac{\p F_\pm^{\delta}}{\p\theta} = 0$, which implies that for an arbitrary but fixed $\varphi$, there exists full integrals $G_{\pm}(\theta,\delta)$ such that
\begin{equation}
dG_{\pm}(\theta,\delta)=F_\pm^{\theta}\ d\theta+ F_\pm^{\delta}\ d\delta.
\end{equation}
Indeed, we can easily obtain
\begin{equation}
G_\pm(\theta,\delta) =  \frac{1}{2} \theta \pm \frac{1}{2} \left( 1 - \frac{a}{\sqrt{a^2-b^2}}\right) \delta.
\end{equation}
Because $G_\pm(\theta,\delta)$ is {\it not}\ a periodic function of either $\theta$ or $\delta$, a path along which both the Hamiltonian and the metric have returned to their initial forms (hence regarded as a closed path) may yield a nonzero Berry phase. For example, after $\delta$ varies from $0$ to $2\pi$ with both $\theta$ and $\varphi$ fixed, the Berry phase is found to be
\begin{equation}
\gamma^B_{\pm}=G_\pm(\theta,2\pi) - G_\pm(\theta,0) = \pm\left(1 - \frac{a}{\sqrt{a^2-b^2}}\right)\pi.
\label{eqn:Berrydelta}
\end{equation}
Similar considerations apply to the parameter pair $(\varphi,\delta)$ because $\frac{\p F_\pm^{\varphi}}{\p\delta} = \frac{\p F_\pm^{\delta}}{\p\varphi}=0.$ We will not repeat the analogous calculations here.

Just like in conventional QM, cases with time-varying parameter pair $(\theta, \varphi)$ turn out to be more involving and interesting. In that case, $\frac{\p F_\pm^{\varphi}}{\p\theta} \ne \frac{\p F_\pm^{\theta}}{\p\varphi}$ and hence there are no simple full integral solutions for the geometric phase $\gamma_\pm^B$ in (\ref{eqn:Berryphase1}).

To calculate $\gamma_\pm^B$, let us define the following vector potential field,
\begin{eqnarray}
{\bf A}_\pm &\equiv& \frac{\hbar}{e}\left(\frac{F_\pm^\theta}{r}{\bf n}^\theta + \frac{F^\varphi_\pm}{r\sin\theta}{\bf n}^{\varphi}\right),
\label{vectorA}
\end{eqnarray}
where $e$ has the dimension of an electric charge. It is straightforward to verify that
\begin{equation}
F_\pm^\theta\rd\theta + F_\pm^\varphi \rd\varphi = \frac{e}{\hbar}{\bf A}_\pm\cdot \rd{\bf r}.
\end{equation}
One therefore has
\begin{equation}
\gamma^B_\pm =\frac{e}{\hbar}\oint {\bf A}_\pm\cdot \rd{\bf r}.
\label{gammaBform}
\end{equation}

Let us then consider the surface enclosed by a closed adiabatic path on the sphere. If it does not include the north pole, then one finds that the vector potential field ${\bf A}$ is well behaved everywhere on this surface. As a result, one may apply Stokes' theorem and reexpress (\ref{gammaBform}) with the surface integral
\begin{equation}
\gamma^B_\pm = \frac{e}{\hbar}\iint \nabla\times {\bf A}_\pm\cdot \rd{\bf S}.
\end{equation}
Further using $\frac{\partial F_{\pm}^{\theta}}{\partial \varphi}=0$ and the explicit form of the curl operator in a spherical coordinate system, this surface integral reduces to
\begin{eqnarray}
\gamma^B_\pm& =& \mp \frac{1}{2}\frac{a}{\sqrt{a^2-b^2}} \iint \sin\theta\, \rd\theta\rd\varphi \nonumber \\
&=&\mp \frac{1}{2}\frac{a}{\sqrt{a^2-b^2}} \Omega ,
\label{nopole}
\end{eqnarray}
where $\Omega$ is just the solid angle covered by the adiabatic path on the sphere.

On the other hand, if the adiabatic path on the sphere does enclose the north pole, then the vector potential field ${\bf A}$ diverges at $\theta=0$ and hence it is illegitimate to apply Stokes' theorem directly. One simple remedy is to divide the adiabatic path into two parts: The first part is a deformed path circumventing the north pole and the second part is an infinitesimal circle surrounding the north pole. In this manner, one may apply Stokes' theorem again to the first part and directly calculate the line integral for the second part. Without loss of generality we always assume counter-clockwise adiabatic paths. Then the first part contributes to the line integral in (\ref{gammaBform}) by again $\mp \frac{1}{2}\frac{a}{\sqrt{a^2-b^2}} \Omega$, and the second part contributes $2\pi F^\varphi_\pm|_{\theta=0}= \left(1\pm\frac{a}{\sqrt{a^2-b^2}}\right)\pi$. Summing up the two contributions, we find
\begin{equation}
\gamma^B_\pm = \mp \frac{1}{2}\frac{a}{\sqrt{a^2-b^2}} \Omega + \left(1\pm\frac{a}{\sqrt{a^2-b^2}}\right)\pi,
\label{pole}
\end{equation}
if a closed adiabatic path encloses the north pole.

In conventional QM the Berry phase of a two-level system in a rotating field can be interpreted as the flux of a fictitious magnetic field. We may do the same here.  In particular, the above expressions of $\gamma^B_\pm$ can be related to a solid angle $\Omega$, plus a constant term depending upon if the adiabatic path encloses the north pole. This suggests that $\gamma^B_\pm$ may be connected with the flux of a magnetic monopole plus a singular field along the $\theta=0$ direction. Specifically,  by combining (\ref{nopole}) and (\ref{pole}), one obtains the Berry phase in terms of the fictitious magnetic field as shown in (\ref{eqn:fictB1}).

\section{Proper mapping between \cP\cT-symmetric and Hermitian two-level systems}
\label{sec:mapping}

Under the similarity transformation $\eta^{(\pm)}$ in (\ref{eqn:eta}), we have two surprisingly simple Hermitian Hamiltonians
\begin{eqnarray}
h^{(\pm)} &\equiv& \eta^{(\pm)} H_{2\times2} \left(\eta^{(\pm)}\right)^{-1}\nn
&=& \epsilon\one_{2\times2} \pm \sqrt{a^2-b^2}\, {\bf n}^r\cdot{\bm \sigma}\nn
&=&\left(
\begin{array}{cc}
\epsilon & 0\\
0 & \epsilon
\end{array}
\right)\pm\sqrt{a^2-b^2}\left(
\begin{array}{cc}
\cos\theta & \sin\theta\, \e^{-\ri\varphi}\\
\sin\theta\, \e^{\ri\varphi} & -\cos\theta
\end{array}
\right).
\label{eqn:hpm}
\end{eqnarray}
Note that $h^{(+)}\leftrightarrow h^{(-)}$ when $\theta$ is shifted by $\pi$, namely, $h^{(+)}$ and $h^{(-)}$ are actually the same Hamiltonian with a shifted parameter $\theta$. Since the $\pm$ signs in front of $\sqrt{a^2-b^2}$ in (\ref{eqn:hpm}) swap two eigenvalues, all the eigenstates of $h^{(+)}$ are also eigenstates of $h^{(-)}$ with swapped eigenvalues. Namely, the normalized eigenstates of $h^{(\pm)}$ are
\begin{eqnarray}
|\phi_+^{(+)}\>&=& |\phi_-^{(-)}\> = \e^{-\ri\frac{\theta}{2}} \left(
\begin{array}{c}
\cos\frac{\theta}{2}\, \e^{-\ri\varphi}\\
\sin\frac{\theta}{2}
\end{array}\right),\nn
|\phi_-^{(+)}\>&=& |\phi_+^{(-)}\> = \e^{-\ri\frac{\theta}{2}} \left(
\begin{array}{c}
-\sin\frac{\theta}{2}\, \e^{-\ri\varphi}\\
\cos\frac{\theta}{2}
\end{array}\right),
\label{eqn:Hermitiankets}
\end{eqnarray}
where we have made an overall phase convention as in the \cP\cT-symmetric case. The advantage for this choice is that in the  limit of $b\to0$, we have
\begin{equation}
\lim_{b\to0} H_{2\times2} = h^{(+)}\quad {\rm and}\quad \lim_{b\to0} |\psi_\pm\> = |\phi_\pm^{(+)}\>.
\end{equation}

To construct a proper mapping from the known improper mapping  $\eta^{(\pm)}$ by (\ref{eqn:improper}), we solve (\ref{eqn:properU}) for the unitary matrix $U$. It is very difficult to obtain $U$ in the general case. In this appendix, we consider paths that involves changes in just one angular variables, either $\delta$ or $\theta$.

\subsection{Varying $\delta$ only}

Plugging the improper mapping in (\ref{eqn:eta}) into (\ref{eqn:properU}) and keeping in mind that only $\delta$ is a function of time, we get a differential equation for $U$ as a function of $\delta$,
\begin{equation}
\frac{\rd U(\delta)}{\rd\delta}= \ri \zeta^{(\pm)} {\bf n}^r\cdot{\bm \sigma} U(\delta),
\label{eqn:Udelta}
\end{equation}
where $\zeta^{(\pm)}$ is defined in (\ref{eqn:dUphi}). For a shorter notation, sometimes we suppress the superscripts in $\zeta^{(\pm)}$. Equation (\ref{eqn:Udelta}) is easy to solve since the factor in front of $U(\delta)$ on the right-hand side is independent of $\delta$. The solutions are
\begin{eqnarray}
U(\delta) &=& \exp(\ri \zeta\delta {\bf n}^r\cdot{\bm \sigma} )U_0\nn
  &=& \left(
    \begin{array}{cc}
    \cos(\zeta\delta) + \ri \cos\theta\sin(\zeta\delta) &
    \ri \sin\theta\sin(\zeta\delta) \e^{-\ri\varphi}\\
    \ri \sin\theta\sin(\zeta\delta) \e^{\ri\varphi} &
    \cos(\zeta\delta) - \ri \cos\theta\sin(\zeta\delta)
    \end{array}
  \right)U_0,
\label{eqn:Udelta2}
\end{eqnarray}
where $U_0$ is an arbitrary constant $2\times2$ unitary matrix.

Remarkably, though the \cP\cT-symmetric Hamiltonian $H_{2\times2}$ is a periodic function of $\delta$ with a period $2\pi$, the found proper mapping, i.e., $U^{\dagger}(\delta)\eta^{(\pm)}$, is {\it not}\ a periodic function of $\delta$ in general. As a consequence, when the parameter $\delta$ varies from $0$ to $2\pi$, the mapped system does not return to itself in general. This explicitly demonstrates that mapping one time-dependent system to another must be carried out carefully.

Another interesting observation can be made. The above matrix structure of $\exp(\ri \zeta\delta {\bf n}^r\cdot{\bm \sigma} )$ is much analogous to that of $h^{(\pm)}$ in (\ref{eqn:hpm}). The consequence is that the two matrices commute with each other. As such, when only $\delta$ varies with time, the properly mapped Hermitian Hamiltonian is simply given by $U_0h^{(\pm)} U^\dag_0$, where $U_0$ is a constant and arbitrary unitary matrix. This being the case, the parameter $\delta$ completely drops out from the mapped Hamiltonian. However, the parameter $\delta$ still appears in the mapped states. For simplicity, let us take $U_0=\one_{2\times2}$ and only consider the mapping to $h^{(+)}$. In this case, the proper mapping $\eta_{\rm proper} ^{(+)} = \left[U^{(+)}(\delta)\right]^\dag\eta^{(+)}$ maps the eigenstates in (\ref{eqn:eigenstates}) to
\begin{equation}
\eta_{\rm proper} ^{(+)}|\psi_\pm\> = \e^{\mp\ri\zeta^{(+)}\delta} |\phi_\pm^{(+)}\>,
\label{eqn:deltakets}
\end{equation}
where $|\phi_\pm^{(+)}\>$ are the conventional choice of the eigenstates for the Hermitian Hamiltonian $h^{(+)}$ in (\ref{eqn:Hermitiankets}). To understand the extra phase factor, $\e^{\mp\ri\zeta^{(+)}\delta}$, let us recall the calculation of the Berry phase in a \cP\cT-symmetric system. As mentioned before, there is always no transition when only $\delta$ varies in time: If we start with an eigenstate of $H_{2\times2}$, the wavefunction [a solution to the Schr\"odinger-like equation (\ref{eqn:PT_schr})] remains as an instantaneous eigenstate. The phase of the wavefunction has two parts, a dynamical phase and a geometric one. For example,
\begin{equation}
|\Psi_\pm\> = \e^{\ri\gamma_\pm^g + \ri \alpha_\pm} |\psi_\pm\>,
\end{equation}
where $\gamma_\pm^g$ is the geometric phase given in (\ref{gammaeq2}) with $\rd\theta=\rd\varphi=0$, $\alpha_\pm$ is the dynamical phase, and $|\psi_\pm\>$ are eigenstates given in (\ref{eqn:eigenstates}). Using the proper mapping, we obtained the mapped time-evolving wave function at the end of the adiabatic process, i.e.,
\begin{equation}
|\Phi_\pm\> \equiv \eta_{\rm proper}^{(+)} |\Psi_\pm\> = \e^{\ri\gamma_\pm^g + \ri \alpha_\pm \mp\ri\zeta^{(+)}\delta} |\phi_\pm ^{(+)}\>.
\end{equation}
Since the mapped Hamiltonian $h^{(+)}$ is time-independent, the phase of the mapped wavefunction must be the dynamical phase and nothing else. Therefore, we get
\begin{equation}
\gamma_\pm^g = \pm\zeta^{(+)}\delta = \pm \frac{1}{2} \left(1-\frac{a}{\sqrt{a^2-b^2}}\right)\delta,
\end{equation}
this is precisely the same geometric phase one would directly get from (\ref{gammaeq2}) if only $\delta$ varies with time. The rather strange Berry phase result in the simple PTQM example here is thus well understood. In general, when $\theta$ or $\varphi$ also changes with time, the proper mapping is much more complicated and cannot be reduced to a simple phase factor.

\subsection{Varying $\theta$ only}
In this case, (\ref{eqn:properU}) reduces to
\begin{eqnarray}
 \frac{\rd U(\theta)}{\rd\theta} &=& \ri \zeta \cos\delta \left(
\begin{array}{cc}
  \sin\theta\sin\delta & (-\cos\theta\sin\delta + \ri\cos\delta )\e^{-\ri\varphi}\\
  (-\cos\theta\sin\delta - \ri\cos\delta )\e^{\ri\varphi} & -\sin\theta\sin\delta
\end{array}
\right)U(\theta)\nn
&=& - \ri \zeta \cos\delta\, \e^{-\ri\frac{\varphi}{2}\sigma_3} \e^{-\ri\frac{\theta}{2}\sigma_2} \e^{\ri\frac{\delta}{2}\sigma_3} \sigma_2 \e^{-\ri\frac{\delta}{2}\sigma_3} \e^{\ri\frac{\theta}{2}\sigma_2} \e^{\ri\frac{\varphi}{2}\sigma_3} U(\theta).
\label{eqn:Utheta}
\end{eqnarray}
This equation cannot be solved by a direct integration because the prefactor of $U(\theta)$ on the right-hand side depends on $\theta$. Noting the factorization of the prefactor, we consider changing variables,
\begin{equation}
V(\theta) \equiv  \e^{\ri\frac{\theta}{2}\sigma_3} \e^{\ri\frac{\varphi}{2}\sigma_2} U(\theta).
\end{equation}
The differential equation satisfied by $V(\theta)$ is
\begin{equation}
\frac{\rd V(\theta)}{\rd\theta} =
\ri\left(- \zeta \cos\delta\, \e^{\ri\frac{\delta}{2}\sigma_3} \sigma_2 \e^{-\ri\frac{\delta}{2}\sigma_3} + \frac{1}{2} \sigma_2  \right)V(\theta).
\end{equation}
This equation can be integrated directly.  In the end, the solutions to the original differential equation in (\ref{eqn:Utheta}) are found to be
\begin{equation}
U(\theta)= \e^{-\ri\frac{\varphi}{2}\sigma_3} \e^{-\ri\frac{\theta}{2}\sigma_2} \exp\left[ \ri \left(- \zeta \cos\delta\, \e^{\ri\frac{\delta}{2}\sigma_3} \sigma_2 \e^{-\ri\frac{\delta}{2}\sigma_3} + \half \sigma_2  \right)\theta\right] U_0.
\end{equation}
From this expression for $U(\theta)$, the obtained proper mapping $U^\dag(\theta)\eta^{(\pm)}$  is {\it not}~a periodic function of $\theta$ in general. Just like in (\ref{eqn:hmapped}), we again use $\Theta$ and $\Phi$ to parameterize the properly mapped Hermitian Hamiltonian,
\begin{equation}
h^{(\pm)}(\theta)=\epsilon\one_{2\times2} \pm \sqrt{a^2-b^2}\left(
\begin{array}{cc}
\cos[\Theta(\theta)] & \sin[\Theta(\theta)]~ \e^{-\ri\Phi(\theta)}\\
\sin[\Theta(\theta)]~ \e^{\ri\Phi(\theta)} & -\cos[\Theta(\theta)]
\end{array}
\right).
\label{eqn:hmapped1}
\end{equation}
For simplicity, we choose $U_0=\one_{2\times2}$. Then the angular parameters for $h^{(\pm)}(\theta)$ are found to satisfy
\begin{eqnarray}
\cos[\Theta(\theta)] &=& \cos\left[\theta\sqrt{1 - 4\zeta\left(1 - \zeta\right) \cos^2\delta}\right], \nn
\e^{\ri\Phi(\theta)} &=& \sqrt{\frac{1-\zeta \left(1+\e^{-2\ri\delta}\right)}{1-\zeta \left(1+\e^{2\ri\delta}\right)}}.
\label{eqn:ThetaPhi}
\end{eqnarray}
Interestingly, this mapped Hermitian Hamiltonian does not depend on $\varphi$. This is accidental due to the arbitrariness in choosing $U_0$. For example, for an alternative choice of the constant unitary matrix given by $U'_0=\e^{\ri \frac{\varphi}{2}\sigma_3}$, $\Theta$ remains the same as in (\ref{eqn:ThetaPhi}) but $\Phi$ is now determined by
\begin{equation}
\e^{\ri\Phi'(\theta)} = \e^{\ri\varphi}\sqrt{\frac{1-\zeta \left(1+\e^{-2\ri\delta}\right)}{1-\zeta \left(1+\e^{2\ri\delta}\right)}}.
\label{eqn:ThetaPhi2}
\end{equation}
To solve for $\Theta$ and $\Phi$ from (\ref{eqn:ThetaPhi}) or (\ref{eqn:ThetaPhi2}), the multi-valued inverse trigonometric functions may be used. In doing so one should be careful in choosing the right branch of the inverse functions.

\end{document}